\documentclass[twocolumn,showpacs,preprintnumbers,amsmath,amssymb]{revtex4}
\usepackage{amssymb}
\usepackage[dvips]{graphicx}
\begin{document}

\title{Diamagnetism of real-space pairs above Tc in hole doped cuprates}

\author{A. S. Alexandrov}

\affiliation{Department of Physics, Loughborough University, Loughborough LE11 3TU, United Kingdom\\
}

\begin{abstract}

The nonlinear normal state diamagnetism reported by Lu Li et al.
[Phys. Rev. B 81, 054510 (2010)] is shown to be incompatible with an
acclaimed Cooper pairing and vortex liquid above the resistive
critical temperature. Instead it is perfectly compatible with the
normal state Landau diamagnetism of real-space composed bosons,
which describes the nonlinear magnetization curves in less
anisotropic cuprates La-Sr-Cu-O (LSCO) and Y-Ba-Cu-O (YBCO) as well
as in strongly anisotropic  bismuth-based cuprates in the whole
range of available magnetic fields.

\end{abstract}

 \maketitle
A growing number of experiments  (\cite{lu,mac,jun,hof,nau,igu,ong}
and references therein) reveal a large diamagnetic response that is
both nonlinear in the magnetic field and strongly T-dependent  well
above the resistive critical temperature T$_c$ of cuprate
superconductors.  The authors of Ref.\cite{lu,ong} suggest that a
long-range phase coherence is destroyed by mobile vortices, however
the off-diagonal order parameter amplitude remains finite and the
Cooper pairing (with a large binding energy) survives up to some
temperature well above T$_c$ supporting a so-called ``preformed
Cooper pair" scenario \cite{kiv}.

Here I show that the anomalous normal state diamagnetism above
resistive T$_c$ reported recently \cite{lu} in crystals of
La$_{2-x}$Sr$_x$CuO$_4$ (LSCO), Bi$_2$ Sr$_{2-y}$ La$_y$CuO$_6$
(Bi2201), Bi$_2$ Sr$_2$CaCu$_2$O$_{8+\delta}$ (Bi2212), and
YBa$_2$Cu$_3$O$_7$ (YBCO) is actually incompatible with the
acclaimed Cooper pairing and vortex liquid, but it is fully
congruent  with the normal state Landau diamagnetism of real-space
composed bosons \cite{aleden,aledia}. The magnetization of less
anisotropic cuprates LSCO and YBCO  is described with the simple
charged-boson magnetization  as well as the magnetization of
quasi-two-dimensional bismuth-based cuprates, described by us
earlier \cite{aledia}.

(i) The extremely sharp resistive transitions measured in high
quality samples at T$_c$ make it impossible to reconcile with the
vortex (or phase fluctuation) scenario as the resistivity looks
perfectly normal showing only moderate magnetoresistance above
T$_c$. Both in-plane \cite{ab} and out-of-plane (see, for example
\cite{c}) resistive transitions remain sharp in the magnetic field
in overdoped, optimally doped and underdoped high quality samples,
providing a reliable determination of the genuine upper critical
field, H$_{c2}$. The sharpness of the transition and little
magnetoresistance argues against the existence of any residual
superconducting order well above T$_c$ (see also \cite{mac}).

(ii) In disagreement with  resistive determinations of the upper
critical field Ref.\cite{lu} claims that H$_{c2}$ is a much higher
field which fully suppresses the diamagnetism. In many cuprates, the
full suppression requires fields as high as 150 Tesla. Such a field
corresponds to a very short  zero temperature in-plane coherence
length, $\xi=\sqrt{\phi_0/2\pi H_{c2}} \lesssim 1.5$nm, which  is
less or about the distance between carriers, $r= \sqrt{2\pi}/k_F$,
in underdoped and overdoped cuprates, respectively ($k_F$ is the
Fermi wave-vector   measured, for example,  in quantum magnetic
oscillation experiments \cite{osc}). The extremely short in-plane
coherence length
 rules out the ``preformed
Cooper pair" scenario, which requires $\xi \gg r$. In cuprates the
pairs do not overlap in underdoped compounds, and they barely touch
at overdoping, so they are not Cooper pairs.

(iii) The authors of Refs. \cite{lu,ong} claim that the profile of
the magnetization $M(B)$ is  consistent with what one would expect
from a vortex liquid in which long-range coherence is destroyed.
This claim is untrue. While the magnitude $|M(B)|$ decreases
logarithmically below T$_c$ as in the conventional vortex liquid,
the set of experimental curves \cite{lu,ong}  show that $|M(B)|$
first \emph{increases} with increasing field $B$ above T$_c$. This
is the opposite of what is expected in the vortex scenario.  This
significant departure from the London liquid behavior is
incompatible with the vortex liquid  above the resistive phase
transition.

(iv)    In the phase-fluctuation scenario \cite{kiv} T$_c$ is
determined by the superfluid density ($x$) rather than by the
density of normal carriers $1+x$. Obviously this scenario  is at
odds with a great number of thermodynamic, kinetic and magnetic
measurements, including recent magnetooscillations \cite{osc}, which
show that only carriers (density $x$) doped into a parent insulator
conduct both in  the normal and superconducting state of
\emph{underdoped} cuprates. On theoretical grounds, the preformed
Cooper-pair scenario contradicts the theorem \cite{leg} that proves
that the number of supercarriers at zero temperature is the same as
the total number of carriers in any clean
 superfluid. The periodic crystal-field potential and electron-electron correlations
 could not change this conclusion. The experimental data \cite{lu} clearly
contradict   the Kosterlitz-Thouless (KT)  scenario of the phase
transition in cuprates, invoked as an origin of the "normal state"
vortex liquid \cite{kiv,ong,lu}. A  magnetization critical exponent
$\delta (T)=\ln B/\ln|M(T,B)|$ for $B\rightarrow 0$, is dramatically
different from the KT universal exponent, $\delta_{KT}(T)=1$, above
T$_c$. While some deviations from  this field-independent KT
critical magnetization have been proposed beyond conventional
scaling \cite{oga}, they could not account for the experimental
$\delta(T)$ and a minimum in $M(T,B)$ observed in high fields above
T$_c$ \cite{lu,ong}. Also the KT critical temperature expressed
through the in-plane penetration depth \cite{kiv}
$k_{B}T_{KT}\approx 0.9d c^2\hbar ^{2}/(16\pi e^{2}\lambda
_{H}^{2})$ appears  several times higher than experimental
values in many cuprates. There are  quite a few samples with about the same %
in-plane penetration depth ${\lambda _{H}}$ and the same inter-plane
distance $d$, but with very different values of T$_{c}$
\cite{alekab}, in disagreement with the KT scenario.

Each inconsistency (i-iv) is individually sufficient to refute the
vortex scenario \cite{lu,ong} of the normal state diamagnetism.
Surprisingly Li Lu \emph{et al.} \cite{lu}  claimed that ``Cooper
pairing is (to their knowledge) the only established electronic
state capable of generating the current response consistent with the
nonlinear, strongly T dependent diamagnetism". These authors
overlooked or neglected our theory of the normal state diamagnetism
 \cite{aleden,aledia}, which quantitatively accounted
for the nonlinear magnetization curves in Bi-2212 (Ref.
\cite{aledia}), and in LSCO, YBCO and Bi2201 as well, as shown here.

Recent quantum Monte Carlo and some other numerical simulations show
that the simplest repulsive Hubbard  model does not explain
high-T$_c$ superconductivity  \cite{imada}. On the other hand, when
a weak to moderate electron-phonon coupling is included, the
superconducting condensation energy is significantly enhanced
\cite{hardy}  and mobile small bipolarons are stabilized
\cite{hague,bonca} as anticipated for strongly correlated electrons
in highly polarizable ionic lattices \cite{dev}. Real-space
tightly-bound pairs, whatever the pairing interaction is, are
described as a charged Bose liquid on a lattice \cite{alebook}. The
superfluid state of such a liquid is the true Bose-Einstein
condensate (BEC), rather than a coherent state of overlapping Cooper
pairs, while the state above T$_c$ is perfectly normal  with no
local or global off-diagonal order.

\begin{figure}
\begin{center}
\includegraphics[angle=-90,width=0.60\textwidth]{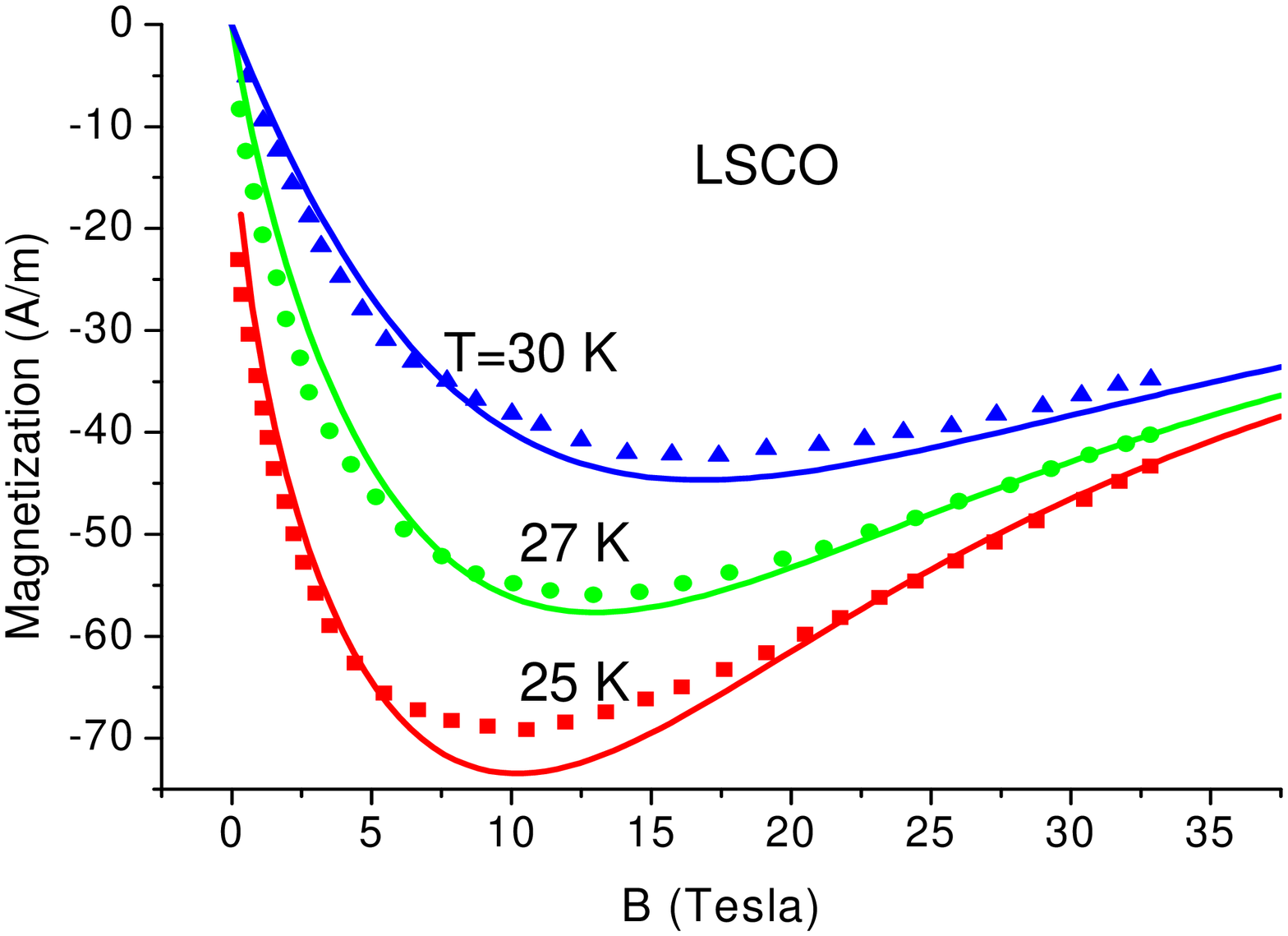}
\includegraphics[angle=-90,width=0.60\textwidth]{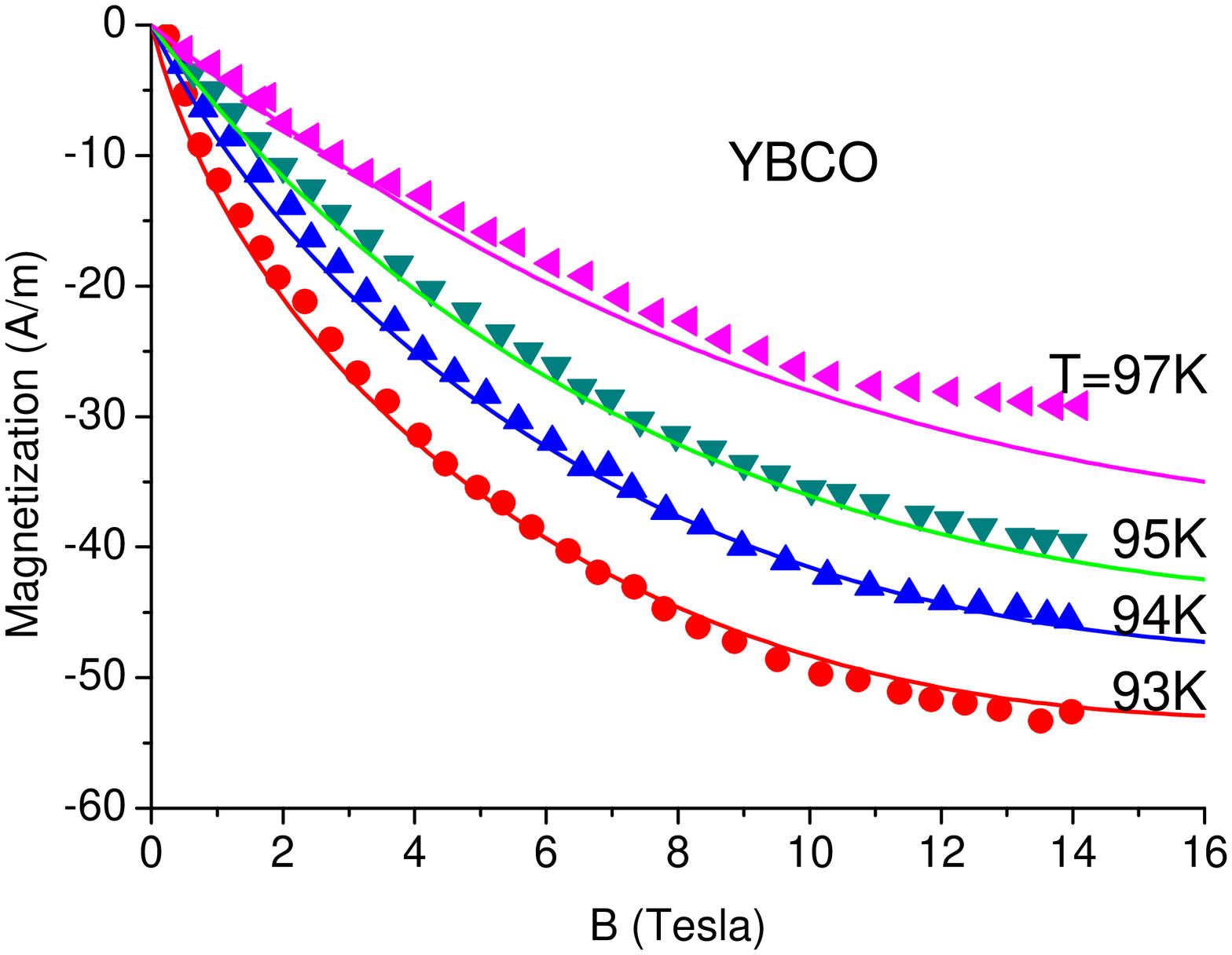}
\vskip -0.5mm \caption{(Color online) Diamagnetism of underdoped
LSCO (symbols, Fig.2 in Ref.\cite{lu}) and optimally doped YBCO
(symbols, Fig.8 in Ref.\cite{lu}) described with Eq.(\ref{M}) above
$T_c$ (lines, T$_c=25$ K, $A=2.75$ A/Tm, $B_1=140$ T,  $B_2=33$ T
for LSCO and T$_c=92$ K, $A=0.45$ A/Tm, $B_1=2300$ T,  $B_2=90$ T
for YBCO).}
\end{center}
\end{figure}

The magnetization of the charged Bose-liquid is given by the simple
expression \cite{aledia}:
\begin{equation}
M(T,B)=-A{B\over {\tau +(B/B_2)^2+\sqrt{B/B_1+\left[\tau
+(B/B_2)^2\right]^2}}}, \label{M}
\end{equation}
which extends  the original Schafroth result \cite{schaf} to the
temperature  region  just  above T$_c$, for $|\tau| = |T/T_c-1|\ll
1$ and $B \ll B_2$. It takes into account the temperature and field
depletion of singlet pairs due to their thermal excitation into spin
triplet and single polaron states split by the magnetic field. The
amplitude $A$ and two characteristic fields, $B_1$ and $B_2$, are
expressed through the zero temperature London penetration depth,
T$_c$ and the spin gap respectively \cite{aledia}. Quite remarkably,
if one fits any of the experimental curves at a certain temperature
$\tau$, all other experimental curves in the applicability  region
of Eq.(\ref{M})  are well described  without any fitting parameters,
as shown for less anisotropic LSCO and YBCO in Fig.1 and for more
anisotropic Bi2201 and Bi2212 in Fig.2. Rather different T$_c$s and
spin gaps account for the significant difference in $B_{1,2}$ for
low-temperature LSCO and Bi2201 compared with high-temperature YBCO
and Bi2212. One can expand  the temperature and magnetic field range
of the theory
 beyond  $|\tau| = |T/T_c-1|\ll 1$ and $B \ll B_2$
solving numerically  exact equations (1,2) of Ref. \cite{aledia} for
the chemical potential and magnetization of charged bosons.
\begin{figure}
\begin{center}
\includegraphics[angle=-90,width=0.60\textwidth]{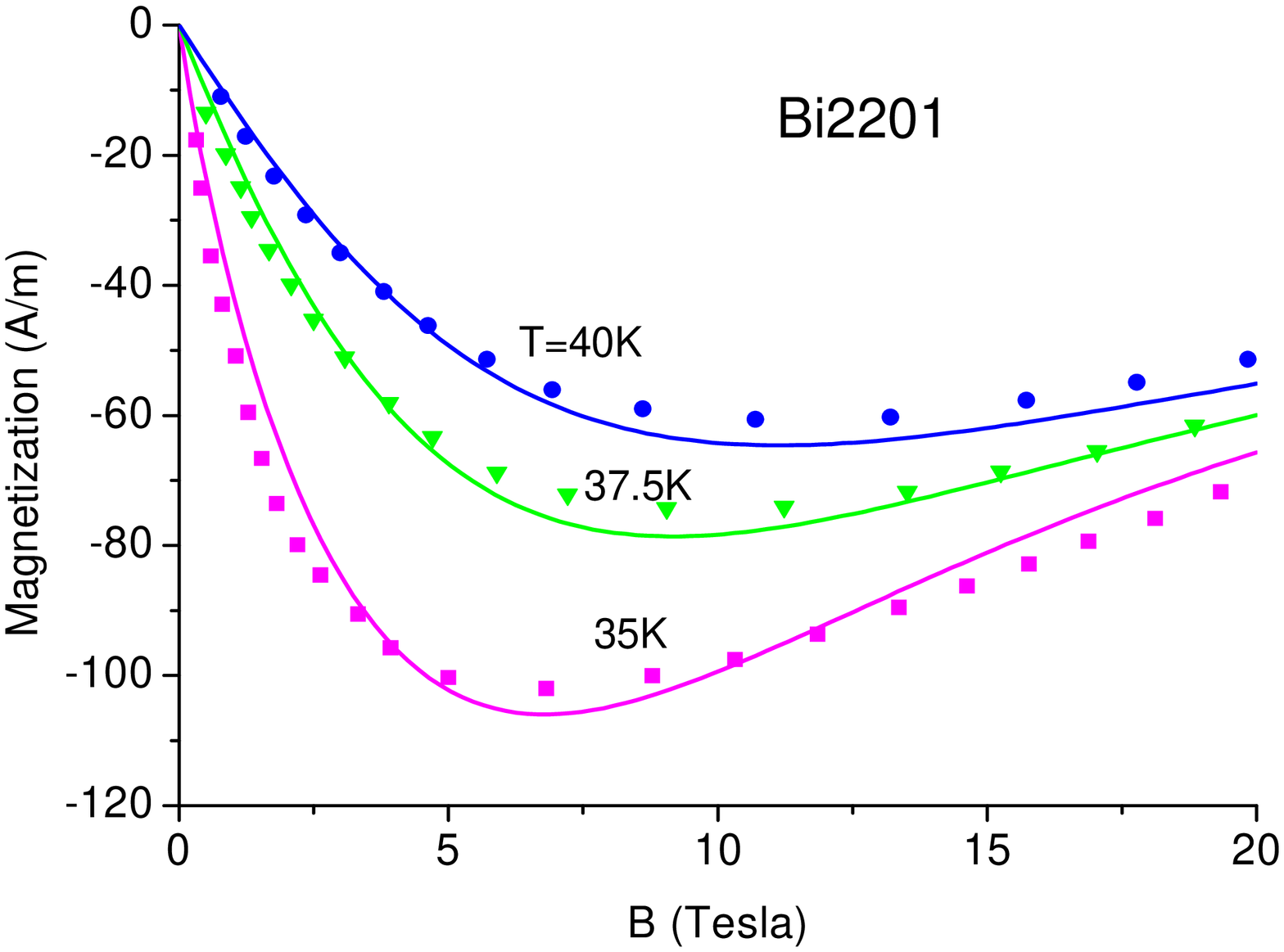}
\includegraphics[angle=-90,width=0.60\textwidth]{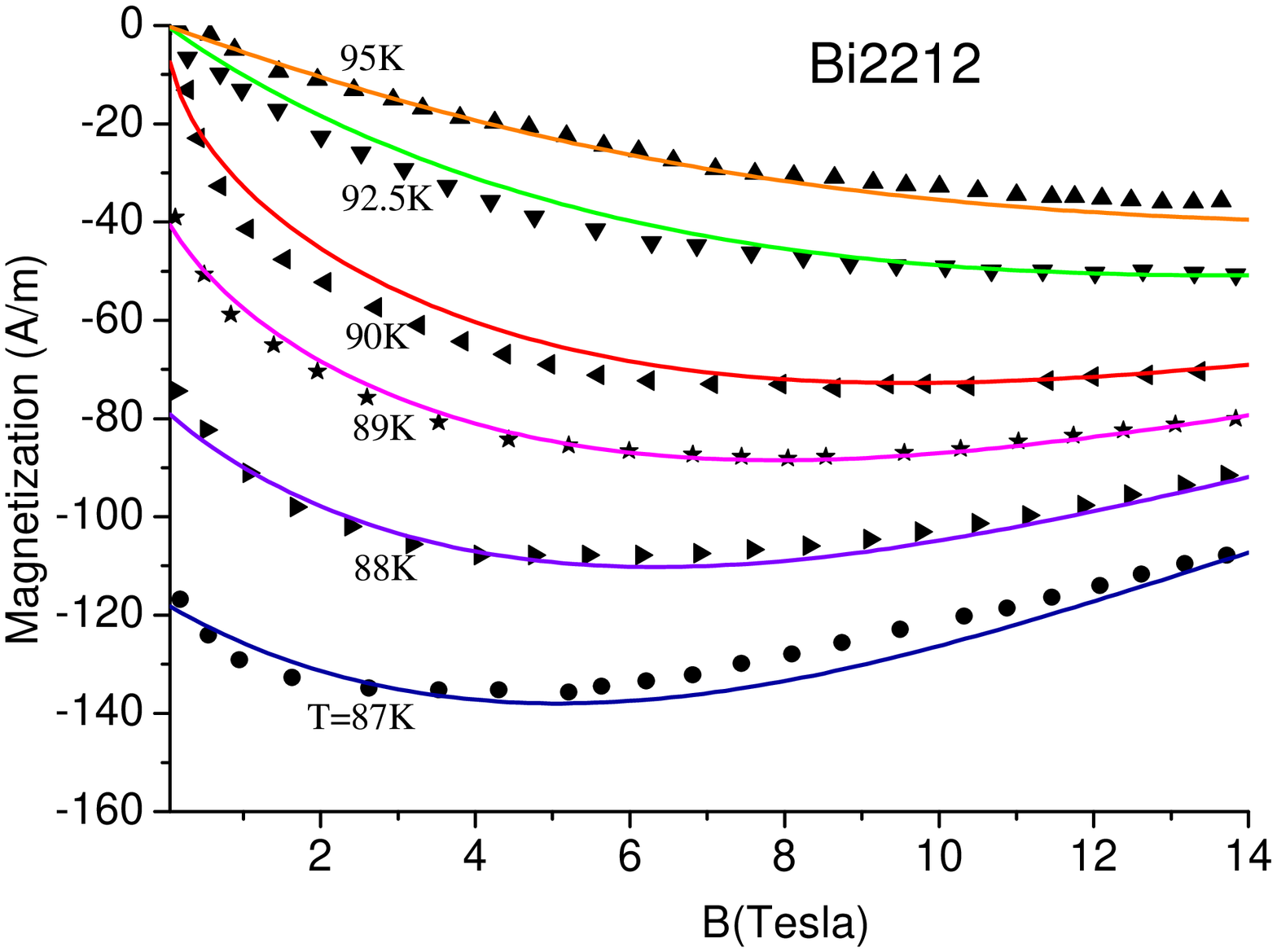}
\vskip -0.5mm \caption{(Color online) Diamagnetism of optimally
doped Bi2201 (symbols, Fig.4c in Ref.\cite{lu}) and optimally doped
Bi2212 (Ref.\cite{aledia}) described with Eq.(\ref{M})  above $T_c$
(lines, T$_c=33.5$ K, $A=4.3$ A/Tm, $B_1=300$T, $B_2=24$ T  for
Bi2201 and T$_c=90$ K, $A=0.62$ A/Tm, $B_1=2832$ T,  $B_2=67$ T for
Bi2212).}
\end{center}
\end{figure}

To conclude I have shown  that the   nonlinear magnetization curves
of quite a few hole-doped cuprates have  a profile characteristic of
normal state real-space composed bosons, rather than "preformed"
Cooper pairs, vortex liquid and the KT phase transition hypothesized
in Refs. \cite{lu,ong,kiv}. There are other independent pieces of
evidence in favor of 3D BEC in cuprate superconductors
\cite{alerev}. The discriminating list includes: a parameter-free
fit of experimental T$_c$ with a BEC T$_{c}$ in a vast number of
cuprates \cite{alekab}; the anomalous Lorentz number pointing to a
double elementary charge per carrier in the normal state \cite{lor};
distinct superconducting and normal state gaps in ARPES, tunnelling,
and pump-probe spectroscopies,  readily explained with real-space
pairs in Refs. \cite{alekim,alebin,mic}, respectively; and  unusual
 upper critical fields \cite{ab,c,ZAV}
expected for charged bosons \cite{aleH}. Importantly, the large
Nernst signal, allegedly supporting vortex liquid in the normal
state of cuprates \cite{lu,ong}, has been explained as the normal
state phenomenon owing to a broken electron-hole symmetry in the
random potential \cite{alezav}, and/or as a result of Fermi-surface
reconstructions \cite{FS}.

I greatly appreciate helpful comments from Joanne Beanland and
Viktor Kabanov.

\end{document}